\documentclass[12pt,reqno]{amsart}
\usepackage{amsfonts,amsmath,amsthm,array,amssymb, amscd}
\usepackage{cite, graphicx, geometry}
\usepackage{url, units}
 \usepackage{hyperref}
\usepackage{epstopdf}
\usepackage{epsfig}
\usepackage{centernot}
\usepackage{multicol}
\makeatletter
\newcommand\appendix@section[1]{%
  \refstepcounter{section}%
  \orig@section*{Appendix \@Alph\c@section: #1}%
  \addcontentsline{toc}{section}{Appendix \@Alph\c@section: #1}%
}
\let\orig@section\section
\usepackage{color}
\definecolor{red}{rgb}{1,0,0}

\definecolor{blue}{rgb}{0,0,1}

\definecolor{gray}{rgb}{0.7,0.7,0.7}

\definecolor{green}{rgb}{0,1,0}

\usepackage[mathlines]{lineno}
\usepackage{mma}
\usepackage[lofdepth,lotdepth]{subfig}
\makeatother

\providecommand{\U}[1]{\protect\rule{.1in}{.1in}}
\providecommand{\U}[1]{\protect\rule{.1in}{.1in}}
\allowdisplaybreaks

\theoremstyle{plain}

\numberwithin{equation}{section}

\input{tcilatex}

\begin{document}
\title[Hidden Symmetry of Harmonic Oscillators]{On a Hidden Symmetry of
Quantum Harmonic Oscillators}
\author{Raquel M. L\'{o}pez}
\address{Mathematical, Computational and Modeling Sciences Center, Arizona
State University, Tempe, AZ 85287--1904, U.S.A.}
\email{rlopez14@asu.edu}
\author{Sergei K. Suslov}
\address{School of Mathematical and Statistical Sciences \& Mathematical,
Computational and Modeling Sciences Center, Arizona State University, Tempe,
AZ 85287--1804, U.S.A.}
\email{sks@asu.edu}
\urladdr{http://hahn.la.asu.edu/\symbol{126}suslov/index.html}
\author{Jos\'{e} M. Vega-Guzm\'{a}n}
\address{Mathematical, Computational and Modeling Sciences Center, Arizona
State University, Tempe, AZ 85287--1904, U.S.A.}
\email{jmvega@asu.edu}
\date{\today }
\subjclass{Primary 81Q05, 35C05. Secondary 42A38}
\keywords{Time-dependent Schr\"{o}dinger equation, generalized harmonic
oscillators, Schr\"{o}dinger group, dynamic invariants, coherent and squeezed states,
Heisenberg uncertainty principle.}

\begin{abstract}
We consider a six-parameter family of the square integrable wave functions
for the simple harmonic oscillator, which cannot be obtained by the standard
separation of variables. They are given by the action of the corresponding
maximal kinematical invariance group on the standard solutions.
In addition, the phase space oscillations of the electron position and linear momentum probability distributions
are computer animated and some possible applications are briefly discussed.
A visualization of the Heisenberg Uncertainty Principle is presented.
\end{abstract}

\maketitle

The purpose of this Letter is to elaborate on a \textquotedblleft
missing\textquotedblright\ class of solutions to the time-dependent Schr\"{o}dinger equation for the simple harmonic oscillator in one dimension.
We also provide an interesting computer-animated feature of these
solutions --- the phase space oscillations of the electron density and
the corresponding probability distribution of the particle linear momentum.
As a result, a dynamic visualization of the fundamental Heisenberg Uncertainty Principle \cite{HeisenbergQM}
is given \cite{Lop:Sus:VegaMath}, \cite{SuslovMath}.

\section{Symmetry and Hidden Solutions}
The time-dependent Schr\"{o}dinger equation for the simple harmonic
oscillator,%
\begin{equation}
2i\psi _{t}+\psi _{xx}-x^{2}\psi =0,  \label{Schroudinger}
\end{equation}%
has the following six-parameter family of square integrable solutions%
\begin{equation}
\psi _{n}\left( x,t\right) =\frac{e^{i\left( \alpha \left( t\right)
x^{2}+\delta \left( t\right) x+\kappa \left( t\right) \right) +i\left(
2n+1\right) \gamma \left( t\right) }}{\sqrt{2^{n}n!\mu \left( t\right) \sqrt{%
\pi }}}\ e^{-\left( \beta \left( t\right) x+\varepsilon \left( t\right)
\right) ^{2}/2}\ H_{n}\left( \beta \left( t\right) x+\varepsilon \left(
t\right) \right) ,  \label{WaveFunction}
\end{equation}%
where $H_{n}\left( x\right) $ are the Hermite polynomials \cite{Ni:Su:Uv} and%
\begin{eqnarray}
\mu \left( t\right) &=&\mu _{0}\sqrt{\beta _{0}^{4}\sin ^{2}t+\left( 2\alpha
_{0}\sin t+\cos t\right) ^{2}},  \label{hhM} \\
\alpha \left( t\right) &=&\frac{\alpha _{0}\cos 2t+\sin 2t\ \left( \beta
_{0}^{4}+4\alpha _{0}^{2}-1\right) /4}{\beta _{0}^{4}\sin ^{2}t+\left(
2\alpha _{0}\sin t+\cos t\right) ^{2}},  \label{hhA} \\
\beta \left( t\right) &=&\frac{\beta _{0}}{\sqrt{\beta _{0}^{4}\sin
^{2}t+\left( 2\alpha _{0}\sin t+\cos t\right) ^{2}}},  \label{hhB} \\
\gamma \left( t\right) &=&\gamma _{0}-\frac{1}{2}\arctan \frac{\beta
_{0}^{2}\sin t}{2\alpha _{0}\sin t+\cos t},  \label{hhG} \\
\delta \left( t\right) &=&\frac{\delta _{0}\left( 2\alpha _{0}\sin t+\cos
t\right) +\varepsilon _{0}\beta _{0}^{3}\sin t}{\beta _{0}^{4}\sin
^{2}t+\left( 2\alpha _{0}\sin t+\cos t\right) ^{2}},  \label{hhD} \\
\varepsilon \left( t\right) &=&\frac{\varepsilon _{0}\left( 2\alpha _{0}\sin
t+\cos t\right) -\beta _{0}\delta _{0}\sin t}{\sqrt{\beta _{0}^{4}\sin
^{2}t+\left( 2\alpha _{0}\sin t+\cos t\right) ^{2}}},  \label{hhE} \\
\kappa \left( t\right) &=&\kappa _{0}+\sin ^{2}t\ \frac{\varepsilon
_{0}\beta _{0}^{2}\left( \alpha _{0}\varepsilon _{0}-\beta _{0}\delta
_{0}\right) -\alpha _{0}\delta _{0}^{2}}{\beta _{0}^{4}\sin ^{2}t+\left(
2\alpha _{0}\sin t+\cos t\right) ^{2}}  \label{hhK} \\
&&+\frac{1}{4}\sin 2t\ \frac{\varepsilon _{0}^{2}\beta _{0}^{2}-\delta
_{0}^{2}}{\beta _{0}^{4}\sin ^{2}t+\left( 2\alpha _{0}\sin t+\cos t\right)
^{2}} . \notag
\end{eqnarray}%
(Here, $\mu _{0}>0,$ $\alpha _{0},$ $\beta _{0}\neq 0,$ $\gamma _{0},$ $\delta
_{0},$ $\varepsilon _{0},$ $\kappa _{0}$ are real initial data.) These
\textquotedblleft missing\textquotedblright\ solutions\ can be derived
analytically in a unified approach to generalized harmonic oscillators (see,
for example, \cite{Cor-Sot:Lop:Sua:Sus}, \cite{Cor-Sot:Sua:SusInv}, \cite%
{Lan:Lop:Sus} and the references therein). They are also verified by a
direct substitution with the aid of \texttt{Mathematica} computer algebra
system \cite{Kouchan11}, \cite{Lop:Sus:VegaMath}, \cite{SuslovMath}.
(The simplest special
case $\mu _{0}=\beta _{0}=1$ and $\alpha _{0}=\gamma _{0}=\delta
_{0}=\varepsilon _{0}=\kappa _{0}=0$ reproduces the textbook solution
obtained by the separation of variables \cite{Flu}, \cite{Gold:Krivch}, \cite{La:Lif}, \cite%
{Merz}; see also the original Schr\"{o}dinger paper \cite{Schroedinger}; and %
the shape-preserving oscillator evolutions %
occur when $\alpha_0=0$ and $\beta_0=1.$ %
More details on the derivation of these formulas can be found in
Refs.~\cite{Lop:Sus:VegaGroup} and \cite{Niederer73}; see also the references therein.)

On the other hand, the \textquotedblleft dynamic harmonic oscillator
states\textquotedblright\ (\ref{WaveFunction})--(\ref{hhK}) are
eigenfunctions,%
\begin{equation}
E\left( t\right) \psi _{n}\left( x,t\right) =\left( n+\frac{1}{2}\right)
\psi _{n}\left( x,t\right) ,  \label{EigenValueProblem}
\end{equation}%
of the time-dependent quadratic invariant,%
\begin{eqnarray}
E\left( t\right) &=&\frac{1}{2}\left[ \frac{\left( p-2\alpha x-\delta
\right) ^{2}}{\beta ^{2}}+\left( \beta x+\varepsilon \right) ^{2}\right]
\label{QuadraticInvariant} \\
&=&\frac{1}{2}\left[ \widehat{a}\left( t\right) \widehat{a}^{\dagger }\left(
t\right) +\widehat{a}^{\dagger }\left( t\right) \widehat{a}\left( t\right) %
\right] ,\qquad \frac{d}{dt}\langle E\rangle =0,  \notag
\end{eqnarray}%
with the required operator identity \cite{Dodonov:Man'koFIAN87}, \cite{SanSusVin}:%
\begin{equation}
\frac{\partial E}{\partial t}+i^{-1}\left[ E,H\right] =0,\qquad H=\frac{1}{2}%
\left( p^{2}+x^{2}\right) .  \label{InvariantDer}
\end{equation}%
Here, the time-dependent annihilation $\widehat{a}\left( t\right) $ and
creation $\widehat{a}^{\dagger }\left( t\right) $ operators are explicitly
given by%
\begin{equation}
\widehat{a}\left( t\right) = \frac{1}{\sqrt{2}}\left( \beta x+\varepsilon +i%
\frac{p-2\alpha x-\delta }{\beta }\right) ,  \qquad 
\widehat{a}^{\dagger }\left( t\right) = \frac{1}{\sqrt{2}}\left( \beta
x+\varepsilon -i\frac{p-2\alpha x-\delta }{\beta }\right)  \label{aacross(t)}
\end{equation}
\noindent
with $p=i^{-1}\partial /\partial x$ in terms of our solutions (\ref{hhA})--(%
\ref{hhK}). These operators satisfy the canonical commutation relation,%
\begin{equation}
\widehat{a}\left( t\right) \widehat{a}^{\dagger }\left( t\right) -\widehat{a}%
^{\dagger }\left( t\right) \widehat{a}\left( t\right) =1,
\label{commutatora(t)across(t)}
\end{equation}%
and the oscillator-type spectrum (\ref{EigenValueProblem}) of the dynamic
invariant $E$ can be obtained in a standard way by using the
Heisenberg--Weyl algebra of the rasing and lowering operators (a
\textquotedblleft second quantization\textquotedblright\ \cite{Akh:Ber}, \cite%
{Lewis:Riesen69}, the Fock states):%
\begin{equation}
\widehat{a}\left( t\right) \Psi _{n}\left( x,t\right) =\sqrt{n}\ \Psi
_{n-1}\left( x,t\right) ,\quad \widehat{a}^{\dagger }\left( t\right) \Psi
_{n}\left( x,t\right) =\sqrt{n+1}\ \Psi _{n+1}\left( x,t\right) .
\label{annandcratoperactions}
\end{equation}%
Here,%
\begin{equation}
\psi _{n}\left( x,t\right) =e^{i\left( 2n+1\right) \gamma \left( t\right) }\
\Psi _{n}\left( x,t\right)  \label{WaveInvariantFunctions}
\end{equation}%
is the relation to the wave functions (\ref{WaveFunction}) with %
$\varphi _{n}\left( t\right) =-\left( 2n+1\right) \gamma \left( t\right)$
%
being the nontrivial Lewis phase \cite{Lewis:Riesen69}, ~\cite{SanSusVin}.

This quadratic dynamic invariant and the corresponding creation and
annihilation operators for the generalized harmonic oscillators have been
introduced recently in Ref.~\cite{SanSusVin} (see also \cite%
{Cor-Sot:Sua:SusInv}, \cite{Dodonov:Man'koFIAN87}, \cite{Suslov10} and the references therein for
important special cases).
An application to the electromagnetic-field quantization and a generalization of the coherent states
are discussed in Refs.~\cite{Kr:Sus12} and \cite{Lan:Lop:Sus:Vega}.

The key ingredients, the maximum kinematical invariance groups of the free
particle and harmonic oscillator, were introduced in \cite{AndersonPlus72},
\cite{AndersonII72}, \cite{Hagen72}, \cite{JACKIW80}, \cite{Niederer72} and
\cite{Niederer73} (see also \cite{BoySharpWint}, \cite{KalninsMiller74},
\cite{Miller77}, \cite{Rosen76}, \cite{SuazoSusVega10}, \cite{SuazoSusVega11}%
, \cite{VinetZhedanov2011} and the references therein). We establish a (hidden symmetry revealing)
connection with certain Ermakov-type system which allows us to bypass a
complexity of the traditional Lie algebra approach \cite{Lop:Sus:VegaGroup}
(see \cite{Ermakov}, \cite{Leach:Andrio08} and the references therein
regarding the Ermakov equation).
(A general procedure of obtaining new solutions by acting on any set of given ones
by enveloping algebra of generators of the Heisenberg--Weyl group is described in \cite{Dodonov:Man'koFIAN87}.)
In addition, the maximal invariance group
of the generalized driven harmonic oscillators is shown to be isomorphic to
the Schr\"{o}dinger group of the free particle and the simple harmonic
oscillator \cite{Lop:Sus:VegaGroup}, \cite{Niederer72}, \cite{Niederer73}.

\section{Discussion}
Quantum systems with quadratic Hamiltonians (see, for example, \cite{Ald:Coss:Guerr:Lop-Ru11},
\cite{Berry84}, \cite{Berry85}, \cite%
{Cor-Sot:Sua:SusInv}, \cite{Dod:Mal:Man75}, \cite{Dodonov:Man'koFIAN87}, \cite{Faddeyev69},
\cite{Fey:Hib}, \cite%
{Har:Ben-Ar:Mann11}, \cite{Malkin:Man'ko79}, \cite{Wolf81}, \cite%
{Yeon:Lee:Um:George:Pandey93}, \cite{Yuen76}, \cite{Zhukov99} and the references therein)
have attracted substantial attention over the years because of their great
importance in many advanced quantum problems. Examples are coherent and squeezed states,
uncertainty relations, Berry's phase, quantization of mechanical systems
and Hamiltonian cosmology. More applications include, but are not limited to
charged particle traps and motion in uniform magnetic fields, molecular
spectroscopy and polyatomic molecules in varying external fields, crystals
through which an electron is passing and exciting the oscillator modes, and
other mode interactions with external fields. Quadratic Hamiltonians have
particular applications in quantum electrodynamics because the
electromagnetic field can be represented as a set of generalized driven harmonic
oscillators \cite{Dod:Klim:Nik93}, \cite{Fey:Hib}.

The maximal kinematical\ invariance group of the simple harmonic oscillator
\cite{Niederer73} provides the six-parameter family of solutions, namely (%
\ref{WaveFunction}) and (\ref{hhM})--(\ref{hhK}), for an arbitrary choice of
the initial data (of the corresponding Ermakov-type system \cite{Ermakov},
\cite{Lan:Lop:Sus}, \cite{Leach:Andrio08}, \cite{Lop:Sus:VegaGroup}). These
\textquotedblleft hidden parameters\textquotedblright\ %
usually disappear after
evaluation of matrix elements and cannot be observed from the spectrum. How
to distinguish between these \textquotedblleft new
dynamic\textquotedblright\ and the \textquotedblleft standard
static\textquotedblright\ harmonic oscillator states (and which of them is
realized in a particular measurement) is thus a fundamental problem.

At the same time, the probability density $\left\vert \psi \left( x,t\right)
\right\vert ^{2}$ of the solution (\ref{WaveFunction}) is obviously moving
with time, somewhat contradicting to the standard textbooks \cite{Flu}, \cite{Gold:Krivch},
\cite{La:Lif}, \cite{Merz}, \cite{Schroedinger}, -- an elementary \texttt{Mathematica} simulation
reveals such space oscillations for the simplest \textquotedblleft dynamic
oscillator states\textquotedblright\ \cite{Lop:Sus:VegaGroup}, \cite%
{Lop:Sus:VegaMath} (see Appendix~A for the Mathematica source code).
The same is true for the probability distribution of the particle linear momentum due to
the Heisenberg Uncertainty Principle \cite{HeisenbergQM}.
These effects, quite possibly, can be observed
experimentally, say in Bose condensates, if the nonlinearity of the
Gross--Pitaevskii equation is turned off by the Feshbach resonance \cite%
{CornWieNobel}, \cite{FedKagShlyapWal}, \cite{Kivsh:Alex:Tur01}, \cite%
{Pit:StrinBook}, \cite{SuazoSuslovSol}, \cite{Suslov11}. A more elementary
example is an electron moving in a uniform magnetic field. By slowly
changing the magnetic field, say, from an initially occupied Landau
level with the standard solution \cite{La:Lif}, \cite{Lop:Sus}, one may
continuously follow the initial wave function evolution (with the quadratic
invariant) until the
magnetic field becomes a constant once again (a parametric excitation; see,
for example, \cite{Cor-Sot:Lop:Sua:Sus}, \cite{Dodonov:Man'koFIAN87}, \cite{Lan:Sus}, \cite%
{Malkin:Man'ko79} and the references therein). The terminal state will 
have, in general, the initial conditions that are required for the
\textquotedblleft dynamic harmonic states\textquotedblright\ (\ref%
{WaveFunction})--(\ref{hhK}) and the probability density should oscillate on
the corresponding Landau level just as our solution predicts. However, it is still not clear
how to observe this effect experimentally (but these \textquotedblleft
dynamic harmonic states\textquotedblright\ will have a nontrivial Berry's phase
\cite{Berry84}, \cite{Berry85}, \cite{SanSusVin}, \cite{Suslov12}, \cite{SuslovMath}).

One may imagine other possible applications, for example, in molecular
spectroscopy \cite{Malkin:Man'ko79}, theory of crystals,
quantum optics \cite{Guerr:Lop:Ald:Coss11}, \cite{Yuen76},
and cavity quantum electrodynamics \cite{Dodonov10},  \cite{Dod:Klim:Nik93},
\cite{Fu:Mat:Hat:Kur:Zeil}, \cite{Kr:Sus12}, \cite{You:Nori11}.
We believe in a dynamic character of the nature \cite{Heisenberg}.
All of that puts the consideration of this Letter into a much broader mathematical
and physical context --- This may help better understand some intriguing features of
quantum motion. (Our example shows that the separation of variables
for the time-dependent Schr\"{o}dinger equations may not always give us
the \textquotedblleft whole picture\textquotedblright.)

\noindent \textbf{Acknowledgments.\/} We wish to thank Professor Sir Michael Berry,
Professor Andrew Bremner, Professor Carlos
Castillo-Ch\'{a}vez, Professor Victor V. Dodonov, Professor Martin Engman,
Professor Georgy Th. Guria, Dr. Christoph Koutschan, Dr. Sergey I. Kryuchkov,
Professor Elliott Lieb, Dr.~Francisco F.~L%
\'{o}pez-Ruiz, Professor Alex Mahalov, Professor Vladimir I. Man'ko, Dr.
Benjamin R. Morin, Professor Peter Paule, Dr. Andrey M. Puchkov,
Priv.-Doz. Dr. Andreas Ruffing, Professor Simon Ruijsenaars,
Professor Vladimir M. Shabaev,
Professor Erwin Suazo,
Dr.  Nikolay Tishchenko,
Professor Luc Vinet and Professor Doron Zeilberger for support, valuable discussions and
encouragement. This paper has been initiated during a short visit of one of
the authors (SKS) to The Erwin Schr\"{o}dinger International Institute for
Mathematical Physics and we thank Professor Christian Krattenthaler, Fakult%
\"{a}t f\"{u}r Mathematik, Universit\"{a}t Wien, for his hospitality. This
research is supported in part by the National Science Foundation--Enhancing
the Mathematical Sciences Workforce in the 21st Century (EMSW21), award \#
0838705; the Alfred P. Sloan Foundation--Sloan National Pipeline Program in
the Mathematical and Statistical Sciences, award \# LTR 05/19/09; and the
National Security Agency--Mathematical \& Theoretical Biology
Institute---Research program for Undergraduates; award \# H98230-09-1-0104.
\appendix
\section{{\tt Mathematica source CODE lines}\cite{Lop:Sus:VegaMath}}\label{mma1}
%
\noindent
{\qquad \ \large{Example~1}}\\
%
{\small
\begin{mma}
\Print The following animation is for the dynamic ground state $n=0$, using $\alpha_0=\gamma_0=\varepsilon_0=0$,
\ $\beta_0=2/3$,\ $\delta_0=1$:\\
\vspace{.1in}
\In |Animate| \Bigg[\text{Plot}\Bigg[\left\{\frac{9 \sqrt{2}\ e^{-\frac{72 \left(x-\ \text{Sin}\left[\frac{1}{500} \pi  (-1+T)\right]\right)^2}{97+65\ \text{Cos}\left[\frac{1}{250} \pi (-1+T)\right]}}}{\sqrt{97+65 \ \text{Cos}\left[\frac{1}{250} \pi (-1+T)\right]}},\ e^{-\frac{4 x^2}{9}}\right\},\text{$\{x,-3.5,3.5\}$,\ \text{AxesLabel}\text{\ -$>$}}\\
 \qquad$\{x,(\text{Abs}[\psi ]){}^{\wedge}2\}$, $\text{PlotRange}\to \{0,2.3\}$,
$\text{Filling}\to \{1\to \text{Bottom}\}$, \text{PlotStyle}$\to \{\text{Thick}, \text{Blue}\} \Bigg],\{T,1001\}\Bigg]$\\
\linebreak
\vfill
\eject
\Out \\

\begin{figure}[h!]
\centering
\subfloat[ ][  ]{
\includegraphics[width=0.18\textwidth]{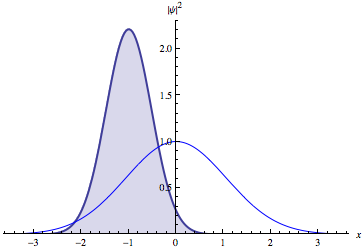}
\label{fig:subfig1a}
}
\subfloat[  ][  ]{
\includegraphics[width=0.18\textwidth]{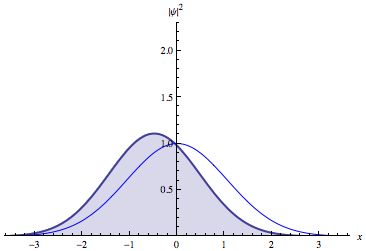}
\label{fig:subfig1b}
}
\subfloat[  ][  ]{
\includegraphics[width=0.18\textwidth]{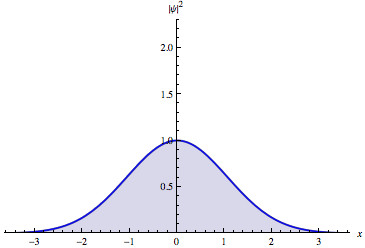}
\label{fig:subfig1c}
}
\subfloat[  ][  ]{
\includegraphics[width=0.18\textwidth]{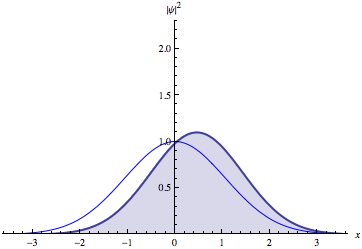}
\label{fig:subfig1d}
}
\subfloat[  ][  ]{
\includegraphics[width=0.18\textwidth]{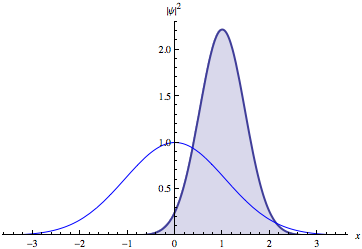}
\label{fig:subfig1e}
}
\caption{Subfigures (a)-(e) are a few {\it stills} taken from the mathematica movie animation \cite{Lop:Sus:VegaMath}.
Starting with (a) they denote the oscillating electron density (blue) of the ground \lq\lq dynamic harmonic state\rq\rq\
of the time-dependent Schr\"odinger equation (\ref{Schroudinger}).  These space oscillations complement (with the help of
{\tt Mathematica})
the corresponding \lq\lq static\rq\rq\ textbook solution (clear)\cite{Schroedinger}.
(The color version of this figure is available only in the electronic edition.)}
\label{fig:globfig2}
\end{figure}
\noindent{\large{Example~2}}\\
\linebreak
\Print The following animation is for the first excited dynamic state $n=1$, using $\alpha_0=\gamma_0=\varepsilon_0=0$,\
$\beta_0=2/3$,\ $\delta_0=1$:\\
\vspace{.1in}
\bigskip
\In |Animate| \Bigg[\text{Plot}\Bigg[\Bigg\{\left(1296 \sqrt{2} \ e^{-\frac{72 \left(x- \ \text{Sin}\left[\frac{1}{500}
\pi (-1+T)\right]\right)^2}{97+65 \ \text{Cos}\left[\frac{1}{250} \pi (-1+T)\right]}} \left(x- \
\text{Sin}\left[\frac{1}{500} \pi (-1+T)\right]\right)^2\right)\\
 \qquad$/\left(97+65\ \text{Cos}\left[\frac{1}{250} \pi (-1+T)\right]\right)^{3/2},\frac{8}{9}\ e^{-\frac{4 x^2}{9}} x^2\Bigg\},\{x,-4.5,4.5\}$, $\text{AxesLabel}\text{\ -$>$}\{x,(\text{Abs}[\psi ]){}^{\wedge}2\}$,\\
\begin{flushright}$\text{PlotRange}\to \{0,1.67\}$, $\text{Filling}\to \{1\to \text{Bottom}\}, \text{PlotStyle}\to \ \{\text{Thick},\text{Blue}\}\Bigg],\{T,1001\}\Bigg]$\\ \end{flushright}
\linebreak
\Out \\
\begin{figure}[h!]
\centering
\subfloat[ ][]{
\includegraphics[width=0.18\textwidth]{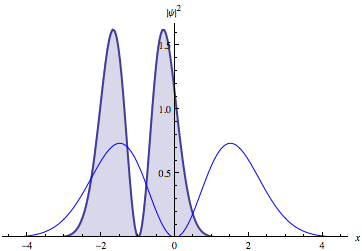}
\label{fig:subfig2a}
}
\subfloat[  ][]{
\includegraphics[width=0.18\textwidth]{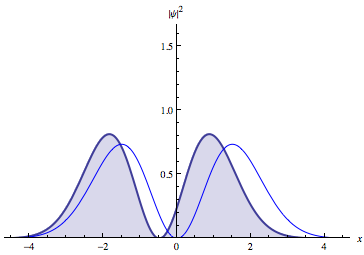}
\label{fig:subfig2b}
}
\subfloat[  ][ ]{
\includegraphics[width=0.18\textwidth]{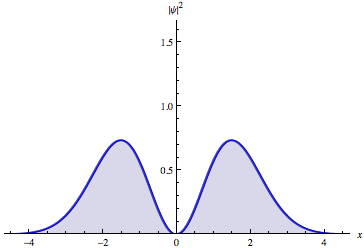}
\label{fig:subfig2c}
}
\subfloat[  ][]{
\includegraphics[width=0.18\textwidth]{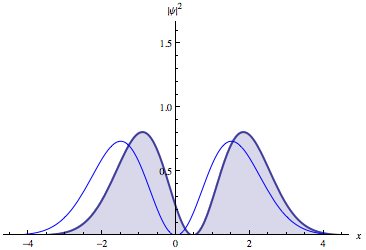}
\label{fig:subfig2d}
}
\subfloat[ ][ ]{
\includegraphics[width=0.18\textwidth]{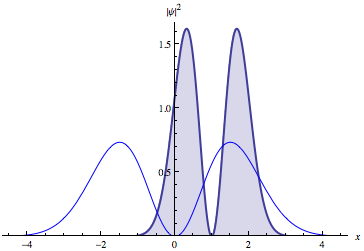}
\label{fig:subfig2e}
}
\caption{Subfigures (a)-(e) are a few {\it stills} taken from the mathematica movie animation \cite{Lop:Sus:VegaMath}.
Starting with (a) they denote the oscillating electron density (blue) of the first excited \lq\lq dynamic harmonic state\rq\rq\
of the time-dependent Schr\"odinger equation (\ref{Schroudinger}).  These space oscillations complement (with the help of {\tt Mathematica})
the corresponding \lq\lq static\rq\rq\ textbook solution (clear)\cite{Schroedinger}.
(The color version of this figure is available only in the electronic edition.)}
\label{fig:globfig2}
\end{figure}
\vfill
\eject
\noindent{\large{Example~3}}\\
\linebreak
\Print The following animations simultaneously show the phase space oscillations of the electron density
and the momentum probability distribution,
according to the Heisenberg Uncertainty Principle, for the dynamic ground state $n=0$ with parameters
$\alpha_0=\gamma_0=\varepsilon_0=\kappa_0=0$, $\beta_0=2/3$ and $\delta_0=3/2$:\\
\vspace{.1in}
\In |Animate| \Bigg[\text{Plot}\Bigg[\Bigg\{
\frac{9 \sqrt{2}\
e^{-\frac{18 \left(2 x+3\ \text{Cos}\left[\frac{1}{500} \pi  (249+T)\right]\right)^2}{97+65 \ \text{Cos}\left[\frac{1}{250} \pi  (-1+T)\right]}
}}%
{\sqrt{97+65 \
\text{Cos}\left[\frac{1}{250} \pi  (-1+T)\right]}} %
\ ,
\\
\vspace{.1in}
\qquad
$
\frac{9 \sqrt{2}\
e^{\frac{18 \left(- 2 x+3 \ \text{Cos}\left[\frac{1}{500} \pi (-1+T)\right]\right)^2
}
{-97+65 \ \text{Cos}\left[\frac{1}{250} \pi  (-1+T)\right]}}}{
\sqrt{97-65\
\text{Cos}\left[\frac{1}{250} \pi (-1+T)\right]}
}\Bigg\}$, 
$\{x,-4.5,4.5\},\text{AxesLabel}
\text{\ -$>$}$ $\{\{x,p\},\{(\text{Abs}[\psi ]){}^{\wedge}2$,\\
\begin{flushright}$(\text{Abs}[a]){}^{\wedge}2\}\},\text{PlotRange}\to \{0,2.3\},\text{Filling}\to \{1\to \text{Bottom}\},\text{PlotStyle}\to \{\text{Thick},\text{Blue}\}\Bigg],\{T,1001\}\Bigg]$\\\end{flushright}
\bigskip
\Out\\
%
\begin{figure}[h!]
\centering
\subfloat[ ][ ]{
\includegraphics[width=0.18\textwidth]{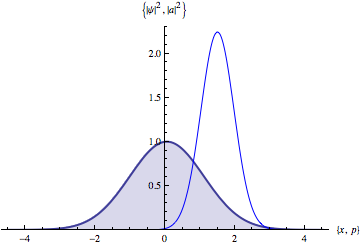}
\label{fig:subfig3g}
}
\subfloat[ ][]{
\includegraphics[width=0.18\textwidth]{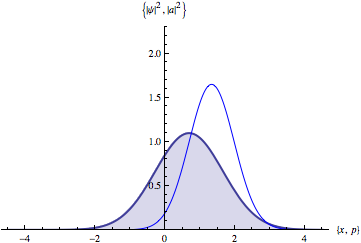}
\label{fig:subfig3a}
}
\subfloat[  ][]{
\includegraphics[width=0.18\textwidth]{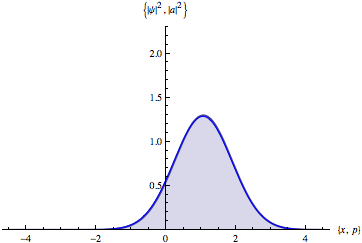}
\label{fig:subfig3b}
}
\subfloat[  ][ ]{
\includegraphics[width=0.18\textwidth]{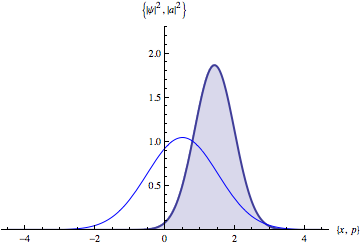}
\label{fig:subfig3c}
}
\subfloat[  ][]{
\includegraphics[width=0.18\textwidth]{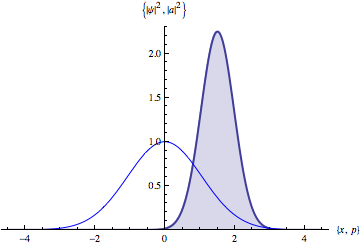}
\label{fig:subfig3d}
}
\\%
\subfloat[ ][ ]{
\includegraphics[width=0.18\textwidth]{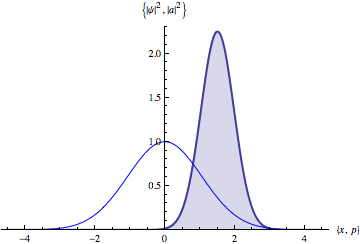}
\label{fig:subfig3e}
}
\subfloat[ ][ ]{
\includegraphics[width=0.18\textwidth]{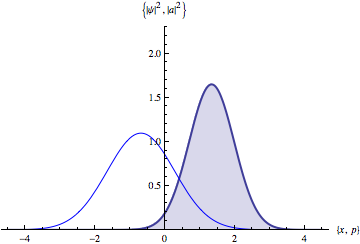}
\label{fig:subfig3f}
}
\subfloat[ ][ ]{
\includegraphics[width=0.18\textwidth]{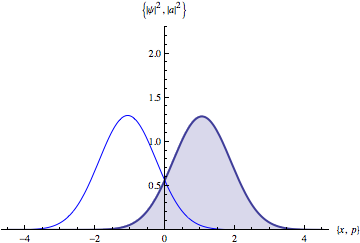}
\label{fig:subfig3g}
}
\subfloat[ ][ ]{
\includegraphics[width=0.18\textwidth]{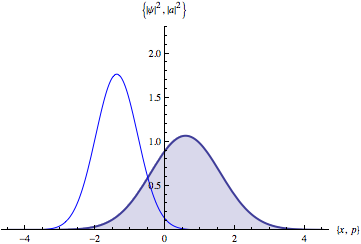}
\label{fig:subfig3h}
}
\subfloat[ ][ ]{
\includegraphics[width=0.18\textwidth]{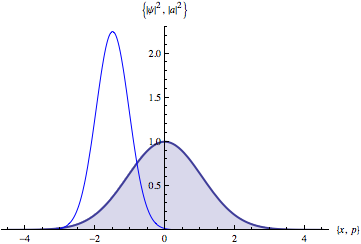}
\label{fig:subfig3g}
}
\caption{Subfigures (a)-(j) are a few {\it stills} taken from the mathematica movie animation \cite{Lop:Sus:VegaMath}.
Starting with (a) they denote simultaneous oscillations of electron density (blue) and probability distribution of momentum (clear) for the ground \lq\lq dynamic harmonic state\rq\rq\ of the time-dependent Schr\"odinger equation (\ref{Schroudinger}).
These phase space oscillations complement (with the help of {\tt Mathematica})
the corresponding \lq\lq static\rq\rq\ textbook solutions \cite{Gold:Krivch}, \cite{Schroedinger}.
(The color version of this figure is available only in the electronic edition.)}
\label{fig:globfig2}
\end{figure}

\end{mma}
}
%

One immediately recognizes from these animations that the particle is the most localized at the turning points
when its linear momentum is the least precisely determined, as required by the fundamental Heisenberg Uncertainty Principle \cite{HeisenbergQM}  --- The more precisely the position is determined, the less precisely the momentum is known in this instant,
and vice versa (see also \cite{UP}). In the creator own words --- \lq\lq If the classical motion of the system is periodic, it may happen that the size of the wave packet at first undergoes only periodic changes\rq\rq\ (see Ref.~\cite{HeisenbergQM}, p.~38).

According to (\ref{aacross(t)}), the corresponding expectation values are
given by%
\begin{eqnarray}
&\langle x\rangle =&-\frac{1}{\beta _{0}}\left[ \left( 2\alpha
_{0}\varepsilon _{0}-\beta _{0}\delta _{0}\right) \sin t+\varepsilon
_{0}\cos t\right] ,\qquad \frac{d}{dt}\langle x\rangle =\langle p\rangle ,
\label{<x>} \\
&\langle p\rangle =&-\frac{1}{\beta _{0}}\left[ \left( 2\alpha
_{0}\varepsilon _{0}-\beta _{0}\delta _{0}\right) \cos t-\varepsilon
_{0}\sin t\right] ,\qquad \frac{d}{dt}\langle p\rangle =-\langle x\rangle
\label{<p>}
\end{eqnarray}%
with the initial data $\left. \langle x\rangle \right\vert%
_{t=0}=-\varepsilon _{0}/\beta _{0}$ and $\left. \langle p\rangle
\right\vert _{t=0}=-\left( 2\alpha _{0}\varepsilon _{0}-\beta _{0}\delta
_{0}\right) /\beta _{0}.$ This provides a classical interpretation of
our \lq\lq hidden\rq\rq\ parameters.

The expectation values $\langle x\rangle $ and $%
\langle p\rangle $ satisfy the classical equation for harmonic motion, $%
y^{\prime \prime }+y=0,$  with the total energy%
\begin{equation}
\frac{1}{2}\left[ \langle p\rangle ^{2}+\langle x\rangle ^{2}\right] =\frac{%
\left( 2\alpha _{0}\varepsilon _{0}-\beta _{0}\delta _{0}\right)
^{2}+\varepsilon _{0}^{2}}{2\beta _{0}^{2}}
=\left. \frac{1}{2}\left[ \langle p\rangle ^{2}+\langle x\rangle ^{2}\right] \right\vert%
_{t=0}
.\label{ClassMechEnergy}
\end{equation}
For the standard deviations,%
\begin{eqnarray}
&\langle \left( \Delta p\right) ^{2}\rangle  =\langle p^{2}\rangle -\langle
p\rangle ^{2}=\left( n+\frac{1}{2}\right)
\dfrac{1+4\alpha _{0}^{2}+\beta%
_{0}^{4}+\left( 4\alpha _{0}^{2}+\beta _{0}^{4}-1\right) \cos 2t-4\alpha
_{0}\sin 2t}{2\beta _{0}^{2}}
,\label{DeltaP} \\
&\langle \left( \Delta x\right) ^{2}\rangle  =\langle x^{2}\rangle -\langle
x\rangle ^{2}=\left( n+\frac{1}{2}\right) \dfrac{1+4\alpha _{0}^{2}+\beta
_{0}^{4}-\left( 4\alpha _{0}^{2}+\beta _{0}^{4}-1\right) \cos 2t+4\alpha
_{0}\sin 2t}{2\beta _{0}^{2}},\label{DeltaX}
\end{eqnarray}%
one gets%
\begin{equation}
\langle \left( \Delta p\right) ^{2}\rangle \langle \left( \Delta x\right)
^{2}\rangle =\left( n+\frac{1}{2}\right) ^{2}\frac{1}{4\beta _{0}^{4}}\left[
\left( 1+4\alpha _{0}^{2}+\beta _{0}^{4}\right) ^{2}-\left( \left( 4\alpha
_{0}^{2}+\beta _{0}^{4}-1\right) \cos 2t-4\alpha _{0}\sin 2t\right) ^{2}%
\right] .\label{HUR}
\end{equation}%
In the case of the Schr\"{o}dinger solution \cite{Schroedinger}, when $%
\alpha _{0}=\delta _{0}=\varepsilon _{0}=0$ and $\beta _{0}=1,$ we arrive at
$\langle x\rangle =\langle p\rangle \equiv 0$ and%
\begin{equation}
\langle \left( \Delta p\right) ^{2}\rangle =\langle \left( \Delta x\right)
^{2}\rangle =n+\frac{1}{2}
\end{equation}
as presented in the textbooks \cite{Flu}, \cite{Gold:Krivch}, \cite{Guerr:Lop:Ald:Coss11},
\cite{Henry:Glotzer88}, \cite{La:Lif}, \cite{Merz}. The dependence on the quantum number $n,$
which disappears from the Ehrenfest theorem \cite{Ehrenfest}, \cite{HeisenbergQM},
is coming back at the level of the higher moments of the distribution.

According to (\ref{HUR}),%
\begin{equation}
\langle \left( \Delta p\right) ^{2}\rangle \langle \left( \Delta x\right)
^{2}\rangle =\left( n+\frac{1}{2}\right) ^{2}\frac{1-4\alpha _{0}^{2}\sin
^{2}2t}{\beta _{0}^{4}},  \label{MUSS}
\end{equation}%
provided that $4\alpha _{0}^{2}+\beta _{0}^{4}=1,$ and the product is equal to
$1/4,$ if $n=0$ and $\sin ^{2}2t=1.$ These are conditions for the
minimum-uncertainty squeezed states of the simple harmonic oscillator
(see, for example, \cite{Henry:Glotzer88}, \cite{Kryuch:Sus:Vega12}).
For the coherent states $\alpha_{0}=0$ and $\beta_{0}=1,$
which describes a two-parameter family
with the initial data $\left. \langle x\rangle \right\vert%
_{t=0}=-\varepsilon _{0}$ and $\left. \langle p\rangle
\right\vert _{t=0}=\delta_{0}.$

The corresponding wave functions in the momentum representation are derived
by the (inverse) Fourier transform of our solutions (\ref{WaveFunction}) and
(\ref{hhM})--(\ref{hhK}) in Appendix~B.
Moreover, in Appendix~C, we explicitly present the action of the Schr\"{o}dinger group
on the wave functions of harmonic oscillators and elaborate on the corresponding
eigenfunction expansion for the sake of \lq completeness\rq.

More examples are available in the authors\rq\ websites.

\section{The Momentum Representation}

For the wave functions in the momentum representation,
\begin{equation}
a_{n}\left( p,t\right) =\frac{1}{\sqrt{2\pi }}\int_{-\infty }^{\infty
}e^{-ipx}\psi _{n}\left( x,t\right) \ dx ,\label{Fourier}
\end{equation}
the integral evaluation is similar to Ref.~\cite{Lan:Lop:Sus}. As a result,
the functions $a_{n}\left( p,t\right)$ are of the same form (\ref{WaveFunction})--(\ref{hhK}),
if $\psi_n \to a_n$ and $x \to p,$ with the initial data
\begin{eqnarray}
&&\alpha _{1}=-\dfrac{\alpha _{0}}{4\alpha _{0}^{2}+\beta _{0}^{2}} ,\qquad \qquad
\beta _{1}=\dfrac{\beta _{0}}{\sqrt{4\alpha _{0}^{2}+\beta _{0}^{2}}} ,\label%
{AB1} \\
&&\gamma _{1}=\gamma _{0}+\dfrac{1}{2} \func{arccot}\dfrac{\beta
_{0}^{2}}{2\alpha _{0}} , \quad \mu _{1}=\mu _{0} \sqrt{4\alpha _{0}^{2}+\beta _{0}^{2}} ,\label{C1} \\
&&\delta _{1}=\dfrac{2\alpha _{0}\delta _{0}+\beta _{0}^{3}\varepsilon _{0}}{%
4\alpha _{0}^{2}+\beta _{0}^{2}},\qquad \quad \varepsilon _{1}=\dfrac{2\alpha
_{0}\varepsilon _{0}-\beta _{0}\delta _{0}}{\sqrt{4\alpha _{0}^{2}+\beta
_{0}^{2}}} ,\label{DE1} \\
&&\kappa _{1}=\kappa _{0}+\dfrac{\alpha _{0}\left( \beta _{0}^{2}\varepsilon
_{0}^{2}-\delta _{0}^{2}\right) +\beta _{0}^{3}\delta _{0}\varepsilon _{0}}{%
4\alpha _{0}^{2}+\beta _{0}^{2}} . \label{K1}
\end{eqnarray}
The calculation details are left to the reader \cite{SuslovMath} (see, for example, Ref.~\cite{Gold:Krivch} for the classical case).%

\section{The Schr\"{o}dinger Group for Simple Harmonic Oscillators%
}

The following substitution
\begin{equation}
\psi\left( x,t\right) =\frac{e^{i\left( \alpha \left( t\right)
x^{2}+\delta \left( t\right) x+\kappa \left( t\right) \right) }}{\sqrt{\mu
\left( t\right) }}\ \chi \left( \xi ,\tau \right) ,\label%
{SchroedingerOscillator}
\end{equation}%
where relations (\ref{hhM})--(\ref{hhK}) hold, transforms the time-dependent
Schr\"{o}dinger equation (\ref{Schroudinger}) into itself with respect to
the new variables $\xi =\beta \left( t\right) x+\varepsilon \left( t\right) $
and $\tau =-\gamma \left( t\right) $ \cite{Niederer73}
(see also \cite{Lop:Sus:VegaGroup} and the references therein).
A \texttt{Mathematica} verification can be found in Refs.~\cite{Kouchan11} and \cite{SuslovMath}.

The eigenfunction expansion of the \lq\lq dynamic harmonic states\rq\rq \ %
with respect to the standard \lq\lq static\rq\rq
ones can be obtain in an obvious way (see, for example, \cite{Lan:Sus} and \cite{Lop:Sus} for similar integral evaluations,
the details will appear elsewhere).
The corresponding matrix elements define the representation
of the  Schr\"{o}dinger group acting on the oscillator wave functions.
(The structure of the Schr\"{o}dinger group in
two-dimensional space-time as a semidirect product of $SL\left( 2, \mathbb{R}
\right) $ and Weyl $W\left( 1\right) $ groups is discussed, for example, in
Refs.~\cite{BoySharpWint}, \cite{KalninsMiller74} and \cite{Miller77}.)

An explicit time evolution of the (bosonic field) creation and annihilation operators for
the \lq\lq dynamic harmonic (Fock) states\rq\rq (with the embedded hidden Schr\"{o}dinger group symmetry)
can be easily derived from (\ref{aacross(t)}), (\ref{annandcratoperactions})
and (\ref{WaveInvariantFunctions}). Applications to the quantization of electromagnetic
fields are discussed in \cite{Kr:Sus12}.


\end{document}